\documentstyle[12pt]{article}
  
\catcode `\@=11
\@addtoreset{equation}{section}
\def\section{\@startsection {section}{1}{\z@}{-3.5ex plus -1ex minus
     -.2ex}{2.3ex plus .2ex}{\large\bf}}
\def\subsection{\@startsection{subsection}{2}{\z@}{-3.25ex plus -1ex minus 
 -.2ex}{1.5ex plus .2ex}{\normalsize\bf}}

\catcode `\@=12

\font\Eul = eufm7 at 12pt

\newcommand{\be}{\begin{equation}}
\newcommand{\en}{\end{equation}}
\newcommand{\bea}{\begin{eqnarray}}
\newcommand{\ena}{\end{eqnarray}}
\newcommand{\beano}{\begin{eqnarray*}}
\newcommand{\enano}{\end{eqnarray*}}
\newcommand{\bee}{\begin{enumerate}}
\newcommand{\ene}{\end{enumerate}}
\newcommand{\bei}{\begin{itemize}}
\newcommand{\eni}{\end{itemize}}

\newtheorem{theorem}{Theorem}[section]
\newtheorem{coroll}[theorem]{Corollary}
\newtheorem{lemma}[theorem]{Lemma} 
\newtheorem{prop}[theorem]{Proposition}  
\newtheorem{defin}[theorem]{Definition}  

\newcommand{\betheo}{\begin{theorem}}
\newcommand{\entheo}{\end{theorem}}

\newcommand{\becor}{\begin{coroll}}
\newcommand{\encor}{\end{coroll}}

\newcommand{\belem}{\begin{lemma}}
\newcommand{\enlem}{\end{lemma}}

\newcommand{\beprop}{\begin{prop}}
\newcommand{\enprop}{\end{prop}}

\newcommand{\beee}{\begin{defin}}
\newcommand{\enee}{\end{defin}}

\newcommand{\proof}{{\em Proof. }-- }
\newcommand{\enproof}{\hfill {$\Box$} \\}

\def\BC{\relax\ifmmode {\!\rlap{\rm C}\hskip 3.2pt 
   {\vrule height 8 pt width 0.5 pt}\,\ }
    \else${\!\rlap{\rm C}\hskip 3.2pt 
    {\vrule height 8 pt width 0.5 pt}\,\ }$\fi}
\def\BR{\relax\ifmmode {{\rm I} \! {\rm R}} \else${{\rm I} \! {\rm R}}$\fi}
\def\BN{\relax\ifmmode {{\rm I} \! {\rm N}} \else${{\rm I} \! {\rm N}} $\fi}

\def\B{\relax\ifmmode {\cal B}\else${\cal B}$\fi}
\def\D{\relax\ifmmode {\cal D}\else${\cal D}$\fi}
\def\H{\relax\ifmmode {\cal H}\else${\cal H}$\fi}
\def\L{\relax\ifmmode {\cal L}\else${\cal L}$\fi}

\def\x{\relax\ifmmode {\mbox{*}}\else*\fi}

\newcommand{\A}{\mbox{\Eul A}}
\newcommand{\gB}{\mbox{\Eul B}}
\newcommand{\N}{\mbox{\Eul N}}

\newcommand{\ha}{^{\rm\textstyle *}}
\newcommand{\ad}{^{\mbox{\scriptsize{\dag}}}}
\newcommand{\sad}{^{\mbox{\tiny{\dag}}}}

\newcommand{\LD}{{\L}\ad(\D,\H)}
\newcommand{\LpD}{{\L}\ad(\D)}
\newcommand{\LDD}{{\L}(\D,\D ')}
\newcommand{\CLDD}{\widetilde{\L}(\D,\D ')}

\newcommand{\xa}{\mbox{*-algebra}}
\newcommand{\qxa}{\mbox{quasi *-algebra}}
\newcommand{\pa}{partial \mbox{*-algebra}}
\newcommand{\po}{partial O\mbox{*-algebra}}

\begin{document}
 
\thispagestyle{empty}

\vspace*{2cm}

\begin{center}
{\large\bf Extension of representations in quasi *-algebras} 
\vspace{2cm}

{\large J.-P. Antoine$^{1}$,
F. Bagarello$^{2}$ and C. Trapani$^{3}$ } \vspace{3mm} \\
$^{1}$ Institut de Physique Th\'eorique, Universit\'e Catholique de Louvain\\
B-1348  Louvain-la-Neuve, Belgium \vspace{2mm}\\ 

$^{2}$ Dipartimento di Matematica dell' Universit\`{a} di Palermo\\ 
I-90123 Palermo, Italy

$^{3}$ Istituto di Fisica dell' Universit\`{a} di Palermo\\ 
I-90123 Palermo, Italy
\end{center}
\vspace{1cm}

\begin{abstract}
Let  $(\A, \A_o)$ be a topological  \qxa, which means in particular that $\A_o$
is a topological  \xa, dense in \A. Let $\pi^o$ be a *-representation of $\A_o$
in some pre-Hilbert space $\D \subset \H$. Then we present several ways of extending 
 $\pi^o$, by closure, to some larger \qxa\ contained in \A, either by Hilbert space operators, 
or by sesquilinear forms on \D. 
Explicit examples are discussed, both abelian and nonabelian, including the CCR algebra.
\vfill

\bigskip
\noindent
E-mail: Antoine@fyma.ucl.ac.be \\ 
\hspace*{13mm} 
Bagarello@ipamat.math.unipa.it 
\\ 
\hspace*{13mm} 
Trapani@ist.fisica.unipa.it 
\bigskip\bigskip

\begin{flushright}
 UCL-IPT-96-18\\ November 1996 
\end{flushright}

\end{abstract}
 
  \newpage
\section{\hspace{-5mm}. Introduction}

One of the most familiar techniques in the description of a quantum system is to put first the
 system in a box $\Lambda$ of finite volume $V$ and then let  $V$ go to infinity, possibly with
suitable boundary conditions (for instance, periodic b. c.). In quantum field theory, this would
correspond to cut-off removals \cite{jafglim}, while in statistical mechanics, the relevant
operation is the thermodynamical limit  \cite{brat}.

A similar approach is common also in the algebraic version of quantum theory. One considers 
first a 
C*-algebra $\A_{\Lambda}$ of observables localized in $\Lambda$ and then let 
$\Lambda \to \infty$. This works in most cases and  leads to the concept of  quasi-local 
observable algebra. However, there are systems for which the technique fails, in the sense that
the dynamics does not converge in a C*-sense. Typical are systems with long range correlations,
for instance the  BCS-Bogoliubov model of a superconductor \cite{thir}-\cite{lass2}, the
description of  the CCR  algebra \cite{lass1,lasst} and various lattice spin systems, called
almost mean field models \cite{bt1,bt2}. In these cases, however, a solution may be found by
taking for the algebra of the full system an O\xa, that is, a \xa\ of unbounded operators on a
fixed invariant domain \cite{schm}, or even a \po\ \cite{ait12,poprev}. The simplest case for
the latter is that of a \qxa\   \cite{lass1,tra1}, and indeed several of the physical systems
listed above lead to such a structure.

First let us recall the basic definitions.
Let \A\ be a vector space and $\A_o$ a  \xa\  contained in \A. 
We say that \A\  is a {\em \qxa\ } with distinguished  \xa\  
$\A_o$ (or, simply, over $\A_o$) if (i) the right and left 
multiplications of an element of \A\ by an element of $\A_o$ are 
always defined and linear; and (ii) an involution * (which 
extends the involution of $\A_o$) is defined in \A\  with the property 
$(AB)\ha = B \ha A\ha$ whenever the multiplication is defined. 
A \qxa\  $(\A, \A_o)$ is said to have a unit $I$ if there 
exists an element $I \in \A_o$ such that $AI =I A =A,\; \forall \, A \in \A$. 
Unless stated otherwise, all the \qxa s used in this paper are assumed to have a unit.
Finally, the \qxa\ $(\A, \A_o)$ is said to be {\em topological} 
if \A\ carries a locally convex topology $\tau$   such 
that (a) the involution is continuous and the multiplications are 
separately continuous; and (b) $\A_o$ is dense in $\A[\tau]$.

Assume now that the set of observables of a given physical system is a \qxa\  
$(\A, \A_o)$. Then  a problem arises. In the standard algebraic formalism, the concrete 
description of the system is obtained by selecting a state $\omega$ on the observable algebra
\A\ and building the corresponding representation by the familiar GNS construction. Even if \A\
is  only a \qxa, the GNS construction is available, as for any \pa\ \cite{poprev,tra1}, but the
notion of state becomes more involved and it is not always obvious to find concrete states. An
alternative approach, possibly easier, would be to proceed in two steps.  (1) Start from the
subalgebra $\A_o$,  select a state $\omega$ over  $\A_o$ and build the corresponding GNS
representation $ \pi^o_\omega $.  (2) Then extend $\pi^o_\omega $ to the full \qxa\ \A, or at
least to a sufficiently large \qxa\  $\A_\pi \subset \A$. The aim of this paper is to explore
this extension process, and more generally, the extension of a given representation $\pi^o$ of
$\A_o$ within the \qxa\ \A. Such an extension naturally proceeds by taking limits, but the
representations $ \pi^o_\omega $ or $\pi^o$ are in general not continuous in the topology of
$\A_o$. Instead, we suggest to perform an  extension by {\em closure}, which requires that we
introduce some notion of closability of the representation.

 We emphasize that this procedure has nothing to do with
the familiar notion of closure of a *-representation $\pi$, which is the extension $\overline{\pi}$
of $\pi$ to the graph topology completion of the domain $\D(\pi)$ (in what follows, we may as well 
assume that $\pi^o$ is a closed \mbox{*-representation} of  $\A_o$). 
The same comment applies to the extension theory developed in \cite{boryng}, which is of a similar
nature. In both cases, the extended set of operators is defined on a larger domain, but remains in
one-to-one correspondence with the original set. Here we want to obtain additional operators.

Let us be more precise.
Let \A\ be a topological \qxa\  over $\A_o$ and $ \pi^o $ be a *-representation of $\A_o$, that 
is,  a map from $\A_o$ into the \xa\ $\LpD$, where \D\ is a dense subspace in some Hilbert space
\H, and $\LpD$ is the set of all  operators $A$ in \H\ such that both $A$ and its adjoint
$A\ha$ map \D\ into itself. In general, extending $ \pi^o $ beyond $\A_o$ will force us to
abandon the invariance of the domain \D. That is, for 
$A \in \A \backslash \A_o$, the extended representative $\pi(A)$ will belong only to  $\LD$, 
which is defined as the set (actually a \po) of all  operators $X$ in \H\ such that
 $D(X)  = \D$ and $D(X\ha)  \supset  \D$. 
Then one may impose on $ \pi^o $ to be closable in $\LD$, and study the corresponding extension 
of $ \pi^o $ by closure. This will be done in Section 2.
In addition, if $ \pi^o $ is the GNS representation $\pi^o_\omega $ associated to some state 
$\omega$ on $\A_o$, there is another possibility of extension, using sesquilinear form 
techniques. We will study this case too in Section 2, and in particular compare the results of
the two methods.

One can also go one step further. Putting on \D\ a suitable (graph) topology, one builds the
 rigged Hilbert space $\D \subset \H  \subset \D'$, where  $\D'$ is the dual of \D\ \cite{gelf},
and observes that
$$
\LpD \; \subset \; \LD \; \subset \; \LDD,
$$
where $\LDD$ denotes the space of all continuous linear maps from \D\ into $\D'$.
Then one may also require that  $ \pi^o $  be closable   in $\LDD$ and try to extend 
$ \pi^o $ within $\LDD$.
This forms the subject matter of Section 3.

First we  observe that, given a \xa\ $\A_o$ and a *-representation 
$ \pi^o : \A_o \to \LpD$, for some prehilbert space $\D \subset \H$, it is possible to build a 
topological \qxa\ $(\A, \A_o)$ contained in $\LDD$ and a *-representation $\pi$ of \A\ that
extends $ \pi^o$. Then we come back to the general problem of extending a *-representation 
$\pi^o$ of $\A_o $ within a {\em given} \qxa\ $(\A, \A_o)$. Since the elements of $\LDD$ may be
interpreted as sesquilinear forms on \D, we will say that the representation $\pi$ is an
{\em extension by  sesquilinear forms}.  Actually this
problem was already addressed in an earlier paper \cite{tra3}, but in a restricted way, in the
sense that only extensions to the whole \qxa\ \A\ were considered.

In both cases, extensions by operators and extensions by sesquilinear forms,
concrete examples will be discussed, abelian ones (\qxa s of functions) as well as nonabelian ones
(\qxa s of operators or matrices). Of course, this paper represents only a first step in the study
of extension of representations. Our aim here is only to identify the problem properly and to
suggest some possible solutions. Further work is in progress.

\section{\hspace{-5mm}. Extensions by Hilbert space operators}

\subsection{\hspace{-5mm}. Closable *-representations in  ${\L}^{\dag }({\D,\H})$}

Let $(\A[\tau], \A_o)$ be a topological \qxa\ and $\pi_o$ a *-representation of $\A_o$ on 
the domain $\D (\pi_o):= \D$.
This means that $\pi_o(A) \in \LpD, \,\, \forall A \in \A_o$.
Since $\LpD \subset \LD$, it make sense to ask the question as to whether $\pi_o$ admits an 
extension to a subspace of $\A$ taking values in $\LD$.

As usual, we consider $\LD$ as endowed with the strong*-topology $t_{\rm s^*}$ defined by the 
family of seminorms
$$ A\in \LD \mapsto \max\{\|Af\|,\|A\ad f\|\}, \quad f \in \D.$$ 
We remind that $\LD$ is $t_{\rm s^*}$-complete.

\begin{defin}
 --
We say that $\pi_o$ is $t_{\rm s^*}$-closable in $\LD$ if, for any net 
$\{X_\alpha\} \subset \A _o $,  
$ X_\alpha \stackrel{\tau}{\longrightarrow} 0 \;\mbox{ and }\; \pi_o(X_\alpha) 
\stackrel{t_{\rm s^*}}{\longrightarrow} Y\in \LD$
imply that  $Y=0$.
\label{main}
\end{defin}

\noindent
If $\pi_o$ is $t_{\rm s^*}$-closable in $\LD$,  we put
$$
\A\ad(\pi)=
\left\{ X \in \A \mid \exists \{X_\alpha \}\subset \A _o: X_\alpha 
\stackrel{\tau}{\longrightarrow} X  {\mbox{ and }} \pi_o(X_\alpha)\;\mbox{is}\; 
t_{\rm s^*}\mbox{-convergent in } \LD \right\}. 
\label{dom0}
$$
For $X \in \A\ad(\pi)$, we set $\pi (X) =  t_{\rm s^*}\mbox{-}\lim\pi_o(X_\alpha)$. Clearly, 
$\pi$ is well-defined and extends $\pi_o$.

Since the involution $\ad$  is $t_{\rm s^*}$-continuous, 
$ X \in \A\ad(\pi)$ implies that $ X\ha \in \A\ad(\pi).$
Therefore $\A\ad(\pi)$ is a *-invariant vector subspace of $\A$, but it need not be a quasi
*-algebra over $\A_o$. Indeed, even if $X \in \A\ad(\pi)$ and $B \in \A_o$, this does not imply
that $XB \in \A\ad(\pi)$. However:

\begin{prop}
--
Let $\pi_o$ be a $t_{\rm s^*}$-closable *-representation of $\A_o$ in $\D$. Assume moreover 
that $\pi_o$ is a bounded *-representation. Then $\A\ad(\pi)$ is a \qxa\ over $\A_o$.
\end{prop}
\proof 
Let $X \in \A\ad(\pi)$ and $B \in \A_o$. We will show that $XB \in \A\ad(\pi)$. Since 
$X \in \A\ad(\pi)$, there exists a net $\{X_\alpha \}\subset \A _o$ such that $ X_\alpha
\stackrel{\tau}{\longrightarrow} X $ and $\pi_o(X_\alpha)$ is $t_{\rm s^*}$-convergent in $\LD$.
Therefore $ X_\alpha B \stackrel{\tau}{\longrightarrow} XB $ and
$$
\|(\pi_o(X_\alpha B)- \pi_o(X_\beta B))f\|=
\|(\pi_o(X_\alpha)- \pi_o(X_\beta ))\pi_o(B)f\|\to 0,
$$
since $\pi_o(B)f\in \D$.
 On the other hand 
\begin{eqnarray*}
\|(\pi_o(X_\alpha B)\ad- \pi_o(X_\beta B)\ad)f\|&=&\|\pi_o(B)\ad (\pi_o(X_\alpha)\ad -
\pi_o(X_\beta )\ad)f\| \\  &\leq& \|\pi_o(B)\| \|(\pi_o(X_\alpha)\ad- \pi_o(X_\beta )\ad)f\|\to
0.
\end{eqnarray*}
Therefore $\pi_o(X_\alpha)B$ is $t_{\rm s^*}$-convergent and so $XB \in \A\ad(\pi)$.
\enproof

\subsection{\hspace{-5mm}. Extension of GNS-representations}

Let $(\A[\tau],\A_o)$ be a topological quasi *-algebra and $\omega$ a state on $\A_o$. 
As is well known \cite{pow}, $\omega$ defines a (closed) *-representation $\pi^o_\omega$ of
$\A_o$ on a domain $\D(\pi^o_\omega)$. We will now try to extend $\pi^o_\omega$ to elements of
$\A$. There are two ways of doing that, which we will discuss below.

The usual GNS-representation $\pi^o_\omega$ is built-up in the following way.
One begins by considering the set
$$ 
\N_\omega = \{ A\in \A_o : \omega(A^*A)=0\},
$$
which turns out to be a left ideal of $\A_o$. Then the quotient space 
$\D_\omega:= \A_o/\N_\omega$ (whose elements will be denoted as $\lambda_\omega(X),$ $X \in
\A_o$) is a pre-Hilbert space with scalar product 
$$
<\lambda_\omega(X),\lambda_\omega(Y)> = \omega(Y^*X), \quad X, Y \in \A_o.
$$
The representation $\pi^o_\omega$ is then defined by
$$
\pi^o_\omega (A)\lambda_\omega(X)=\lambda_\omega(AX).
$$ 
One readily checks that $\pi^o_\omega (A\ha)=\pi^o_\omega (A)\ad,\; \forall \, A \in \A_o$ 
and so $\pi^o_\omega (A) \in {\cal L}\ad (\D_\omega)$. Furthermore, $\lambda_\omega(I)$ is a
cyclic vector for $\pi^o_\omega$. Finally this representation can be continued to a {\em closed}
representation in a standard fashion, but we are not interested here to this procedure.

If $\pi^o_\omega$ is $t_{\rm s^*}$-closable, then we can proceed as before and we obtain an 
extension 
$\pi_\omega$ of $\pi^o_\omega$ to a subspace $\A\ad(\pi_\omega)$ of $\A$. However, there is 
 another possible way of extending $\pi^o_\omega$ outside of $\A_o$, namely by closing the
positive sesquilinear form $\Omega_o$ defined by $\omega$.

The equation
\be 
\Omega_o(A,B)= \omega(B^*A), \quad A,B \in \A_o
\label{4.1}
\en 
defines indeed a positive sesquilinear form on $\A_o \times \A_o$. We now assume that 
$\Omega_o$ is closable in $\A$ in the sense we shall define in a while. Let
$\{X_\alpha\}$ be a net in $\A_o$ and $X\in \A$. We say that $X_\alpha$ $\Omega_o$-converges to
$X$ if  $X_\alpha \stackrel{\tau}{\longrightarrow} X$ and
$\Omega_o(X_\alpha-X_\beta,X_\alpha-X_\beta)\rightarrow 0$.

\begin{defin}
--
We say that $\Omega_o$ is closable if $\Omega_o(X_\alpha,X_\alpha)$ converges to $0$ for any net
$\{X_\alpha\}$  that 
$\Omega_o$-converges to $0$, 
.
\end{defin} 
If $\Omega_o$ is closable, we define
\be
\D_{\Omega}=\left\{ X \in \A : \exists \{X_\alpha \}\subset \A _o \;\mbox{ s.t. }\; X_\alpha 
\mbox{ $\Omega_o$-converges to } X\right\}.
\en
For $X,Y \in \D_{\Omega}$, with $X=\lim_{\alpha}X_\alpha$ and  $Y=\lim_{\alpha}Y_\alpha$ we put
$$ 
\Omega (X,Y)= \lim_{\alpha}\Omega_o(X_\alpha,Y_\alpha ).
$$

\vspace{3mm}
{\bf Remark} -- The above definitions are slight modifications of the usual definition of 
closable sesquilinear form on a Hilbert space \cite{kato}.

\vspace{3mm}
Let $ \Gamma^o_{\Omega}:=\D_{\Omega} \times \D_{\Omega}$.
It is easily seen that $\Omega$ and $\Gamma^0_{\Omega}$ satisfy the following conditions 
\cite{weights}:  
\begin{itemize}
\item[(D1)]  
$\Gamma^o_\Omega$ preserves linearity : if $(X,Y) \in
\Gamma^o_\Omega$  and $(X,Z) \in \Gamma^o_\Omega$, then  $(X,Y + \lambda Z)
 \in \Gamma^o_\Omega,\; \forall \lambda \in {\BC}$;
\item[(D2)]
  $\Gamma^o_\Omega$ is symmetric, i.e. if $(X,Y) \in \Gamma^o_\Omega$, then  
$(Y,X) \in \Gamma^o_\Omega $; 
\item[(D3)]
  $\Omega$ is hermitian, i.e. $\Omega(X,Y)  =
\overline{\Omega(Y,X)},\; \forall (X,Y) \in \Gamma^o_\Omega$; 
\item[(D4)]
  $\Omega$ is positive, i.e. $\Omega(X,X) \geq 0$, $\forall X \in \D_{\Omega}$.
\end{itemize}

Let us now define the set
\be \A_\Omega =\left\{ X \in \D_{\Omega}: X^*\in \D_{\Omega} \mbox{ and }XB \in 
\D_{\Omega}, \;\; \forall B \in \A_o\right\}.  
\en
It is clear that $\A_\Omega$ is a \qxa\ over $\A_o$. The form $\Omega$ may be regarded as an 
{\em everywhere defined} $\A_o$-weight in the sense of \cite[Definition 3.1]{weights} with the
obvious choice $\Gamma_\Omega = \A_\Omega \times \A_\Omega$. In fact the following conditions
hold:
\begin{itemize}
\item[(i)]
$ \A_o \times  \A_o \subseteq \Gamma_\Omega$;
\item[(ii)]
  If $X \in \A_\Omega $ and $B \in \A_o $, then $(XB, C) \in \Gamma_\Omega, \forall \,
C\in \A_o $;
\item[(iii)] 
 $\Omega(XB_1,B_2)  =  \Omega(B_1 ,X^* B_2), \forall \, X\in \A_\Omega, B_1,B_2 \in
\A_o $;
 
\item[(iv)] If $\Omega(X,X)=0$  for some $X\in \A_\Omega$,
then $\Omega(X,Y)=0, \; \forall \, Y \in \A_\Omega.$
\end{itemize}

Indeed, (i) and (ii) are more or less obvious; (iii) follows from a simple limiting argument
 and (iv) is due to the fact that 
$\Omega$ is a positive sesquilinear form on $\A_\Omega \times \A_\Omega$ and thus satisfies the
 Cauchy-Schwarz inequality
$$
|\Omega(X,Y)|^2 \leq \Omega(X,X)\Omega(Y,Y), \quad \forall X,Y \in \A_\Omega.
$$
In conclusion, the propositions given in \cite{weights} can be applied. We obtain a GNS 
representation $\pi_\Omega$ in a partial inner product space and $\pi_\Omega$ is an extension of
$\pi^o_\omega$.

Conditions (i)-(iii) are characteristic of ips (invariant positive sesquilinear) forms on
 partial *-algebras \cite{ait12,poprev}. (The complete definition of ips form on a partial
*-algebra actually includes an additional condition, but in this case it follows immediately
from (iii).) Nevertheless, the theory of *-representations developed there can be applied only
if an additional density condition is fulfilled. This is actually the case, as we shall see in a
while.

One begins with considering the set
$$
\N_\Omega = \{ A\in \A_\Omega : \Omega(A,A)=0\}
$$
and then takes the quotient $\A_\Omega /\N_\Omega =: \lambda_\Omega (\A_\Omega)$ whose elements 
are denoted as $\lambda_\Omega(A)$, $A \in \A_\Omega$. Let $\H_\Omega$ be the completion of
$\lambda_\Omega (\A_\Omega)$. Let $\lambda_\Omega (\A_o)= \{\lambda_\Omega(B)\,,\,B\in \A_o\}$.
Then $\lambda_\Omega (\A_o)$ is dense in $\H_\Omega$. Indeed, $\A_\Omega \subseteq
\D_{\Omega}$. Therefore, by the construction itself, if $\lambda_\Omega(A) \in \lambda_\Omega
(\A_\Omega)$, there exists a net $\{\lambda_\Omega(A_\alpha)\} \subset \lambda_\Omega (\A_o)$
which converges to $\lambda_\Omega(A)$ with respect to the norm of $\H_\Omega$.
 Then $\Omega$ is an ips form on $\A_\Omega$ (see the Remark above) and thus a GNS construction
 can be performed as in \cite{ait12} ; the operators obtained in this way live in the Hilbert
space $\H_\Omega$.  More precisely, one defines, for $A \in \A_\Omega$:
$$ 
\pi_\Omega (A)\lambda_\Omega(B) = \lambda_\Omega(AB), \quad B \in \A_o.
$$   
Then $ \pi_\Omega (A)$ is a well-defined linear operator from $\lambda_\Omega(\A_o)$ into 
$\H_\Omega$. From (iii) it follows that, for each $A \in \A_\Omega$, $ \pi_\Omega (A) \in
{\cal L}\ad(\lambda_\Omega(\A_o),\H_\Omega )$. In particular, if $A \in \A_o$ then
$ \pi_\Omega (A)$ maps $\lambda_\Omega (\A_o)$ into itself.

The next step, of course, is to compare the results obtained in the two ways explained above.
The first natural question is whether the closability of $\Omega_o$ implies, or is implied by, 
the closability of $\pi^o_\omega$.
The second question is, what kind of relation exists, if any, between $\A\ad(\pi_\omega)$ and 
$\A_\Omega$.

We begin by a subsidiary result that will be needed later.
 First, we define
$$\Omega_o^*(X,Y) = \Omega_o(Y^*,X^*), \quad X,Y \in \A_o.
$$
Furthermore, for $B \in \A_o$, we set
$$
\Omega^o_B(X,Y) = \Omega_o(XB,YB), \quad X,Y \in \A_o.
$$
The forms $\Omega_o^*$ and $\Omega^o_B,\;B \in \A_o$, are still positive sesquilinear forms on 
$\A_o\times \A_o$.
Then one has:

\begin{lemma} -- Let $\Omega_o$ a positive sesquilinear form on $\A_o\times \A_o$. 
The following statements are equivalent
\begin{itemize}
\item[(i)]$\Omega_o$ is closable;
\item[(ii)]$\Omega_o^*$ is closable;
\item[(iii)]$\Omega^o_B$ is closable, for each $B \in \A_o$.
\end{itemize}
\label{lemma}
\end{lemma}

\proof 
\underline{(i) $\Rightarrow$ (ii)}: 
Let $\Omega_o$ be closable and let $\{X_\alpha\}$ be a net in $\A_o$  that
 $\Omega_o\ha$-converges to $0$. Then $X_\alpha \stackrel{\tau}{\longrightarrow} 0$ 
and $\Omega_o(X\ha_\alpha-X\ha_\beta,X\ha_\alpha-X\ha_\beta) \rightarrow 0$. Since also $X\ha_\alpha
\stackrel{\tau}{\longrightarrow} 0$, we get that $X\ha_\alpha$ is $\Omega_o$-convergent to $0$;
thus $\Omega_o(X\ha_\alpha,X\ha_\alpha)=\Omega_o\ha(X_\alpha,X_\alpha)\rightarrow 0$.

\noindent 
\underline{(ii) $\Rightarrow$ (i)}: this follows from the previous implication by taking into account
that $(\Omega_o\ha)\ha=\Omega_o$.

\noindent 
\underline{(i) $\Rightarrow$ (iii)}:  Let $X_\alpha$  $\Omega^o_B$-converge to $0$. Then
$X_\alpha \stackrel{\tau}{\longrightarrow} 0$ and
$\Omega_o((X_\alpha-X_\beta)B,(X_\alpha-X_\beta)B) \rightarrow 0$. Since also $X_\alpha B
\stackrel{\tau}{\longrightarrow} 0$, we conclude that $\{X_\alpha B\}$ $\Omega_o$-converges to
$0$. Therefore, by the closability of $\Omega_o$,  $\Omega^o_B(X_\alpha,X_\alpha)\rightarrow 0$.
Hence $\Omega^o_B$ is closable, for each $B \in \A_o$.

\noindent 
\underline{(iii) $\Rightarrow$ (i)}: follows from the fact that $\A_o$ contains the unit.
\enproof

\begin{prop} --
Let $\Omega_o$ be closable. Then $\pi^o_\omega$ is $t_{\rm s^*}$-closable in 
${\cal L}\ad(\D_\omega, \H_\omega)$ and one has $\A\ad(\pi_\omega)\subset\A_\Omega$
\end{prop}

\proof
Let $\{X_\alpha\}$ be a net in $\A_o$ such that $X_\alpha \stackrel{\tau}{\longrightarrow} 0$
 and $\pi^o_\omega (X_\alpha) \stackrel{t_{\rm s^*}}{\longrightarrow} Y \in \LD$ with
$\D=\D_\omega$ and $\H=\H_\omega$. Recalling that elements of $\D_\omega$ are simply cosets
$\lambda_\omega (B),\; B\in \A_o$, we get, in particular
$$ 
\|\pi^o_\omega (X_\alpha-X_\beta)\lambda_\omega (B)\|^2 = \Omega_o((X_\alpha-X_\beta) B,
(X_\alpha-X_\beta) B) \to 0
$$ 
$$ 
\|\pi^o_\omega (X\ha_\alpha-X\ha_\beta)\lambda_\omega (B)\|^2 = \Omega_o((X\ha_\alpha-X\ha_\beta) B, 
(X\ha_\alpha-X\ha_\beta) B) \to 0
$$
and 
$$ 
\|\pi^o_\omega (X_\alpha)\lambda_\omega (B)\|^2 = \Omega_o(X_\alpha B, X_\alpha B) \to \|Y 
\lambda_\omega (B)\|^2.
$$ 
$$ 
\|\pi^o_\omega (X\ha_\alpha)\lambda_\omega (B)\|^2 = \Omega_o(X\ha_\alpha B, X\ha_\alpha B) \to 
\|Y\ad \lambda_\omega (B)\|^2.
$$ 
But by Lemma \ref{lemma}, it follows that $\|Y \lambda_\omega (B)\|=\|Y\ad \lambda_\omega (B)\|=0,
\; \forall \, B\in \A_o$. Then $\pi^o_\omega$ is $t_{\rm s^*}$-closable in ${\cal
L}\ad(\D_\omega, \H_\omega)$.

 Let now $X \in \A\ad(\pi_\omega)$; then, by definition, there exists a net 
 $\{X_\alpha\} \subset \A_o$ such that $X_\alpha \stackrel{\tau}{\longrightarrow} X $ and 
$\pi^o_\omega(X_\alpha) \stackrel{t_{\rm s^*}}{\longrightarrow} \pi_\omega (X)$. This implies
that both $\|\pi^o_\omega(X_\alpha-X_\beta) \lambda_\omega (B)\|$ and
$\|\pi^o_\omega(X\ha_\alpha-X\ha_\beta) \lambda_\omega (B)\|$ tend to $0$ for any $B \in \A_o$. But
$\|\pi^o_\omega(X_\alpha-X_\beta) \lambda_\omega (B)\|^2= \Omega_o((X_\alpha-X_\beta) B,
(X_\alpha-X_\beta) B)$ and $\|\pi^o_\omega(X\ha_\alpha-X\ha_\beta) \lambda_\omega (B)\|^2=
\Omega_o((X\ha_\alpha-X\ha_\beta) B, (X\ha_\alpha-X\ha_\beta) B)$. These equalities imply easily that
$X \in \A_\Omega$.
\enproof

The converse statements may be proven under the much stronger assumption that, if 
$\{X_\alpha\} \subset \A _o $ is a net $\Omega_o$-convergent to $0$, then both
$\Omega^o_B(X_\alpha - X_\beta,X_\alpha - X_\beta)$ and 
$\Omega^o_B(X_\alpha\ha - X_\beta\ha,X_\alpha\ha - X_\beta\ha)$ converge to 0, for all 
$B \in  \A _o$. Then one gets $\A\ad(\pi_\omega) = \A_\Omega$, but this result does not seem very
useful in practice.
\vspace{4mm}

\noindent {\bf Examples 2.6} --
 (1)   Let $X=[0,1]$, $\A=L^p(X), \; p \geq 1$ and $\A_o = C(X)$, the 
C*-algebra of continuous functions on $X$. 
Let $\omega$ be the linear functional on $C(X)$ defined by
$$
 \omega (f) = \int_X f(x) dx, \quad f \in C(X).
$$ 
The sesquilinear form  $\Omega_o$ associated with $\omega$ is then defined as
$$ 
\Omega_o (f,g)= \int_X f(x)\overline{g(x)} dx, \quad f,g \in C(X).
$$
If $p \geq 2$, then $\Omega_o$ is bounded, as it is easily seen, and so it can be extended to 
the whole space $L^p(X)$.

If $1\leq p < 2$, the situation is different. In this case, in fact, $\Omega_o$ is only a 
closable sesquilinear form. 
Indeed, assume that $\|f_n\|_p \to 0$ and that $ \Omega_o
(f_n-f_m,f_n-f_m)= \|f_n-f_m\|_2^2$ is convergent to $0$. Then, there exists an element $f \in
L^2(X)$ such that $\|f_n-f\|_2^2 \to 0$. This implies that $f_n$ converges to $f$ in measure. The
convergence of $f_n$ to $0$ in $L^p(X)$ in turn implies the convergence of $f_n$ to $0$ in
measure. And so $f=0$ a.e. in X. Therefore $\Omega_o(f_n,f_n)=\|f_n\|_2^2 \to 0$.
Hence $\Omega_o$ is closable. It is easy to check that $L^2(X) \subset
\A_\Omega $. To get the opposite inclusion, let us consider $f \in \D_{\Omega}$. Then there
exists a sequence $\{f_n\}$ such that $\|f_n - f\|_p \to 0$ and $\Omega(f,f)= \lim_{n
\rightarrow \infty}\int_X|f_n|^2dx$. The convergence of $f_n$ to $f$ implies the existence of a
subsequence $f_{n_k}$ converging to $f$ a.e. in $X$. Then also $|f_{n_k}|^2$ converges to
$|f|^2$ a.e. in $X$ and $\lim_{n \rightarrow \infty}\int_X|f_n|^2dx$ exists by assumption. By
Fatou's lemma, it follows that $f \in L^2(X)$ and therefore
$\A_\Omega = L^2(X)$. 

In this case, we also have $\A\ad(\pi_\omega) = L^2(X)$. From the previous discussion  and 
from Proposition 2.5, we know that $\A\ad(\pi_\omega) \subset \A_\Omega = L^2(X)$. Let now $f
\in L^2(X)$; so there exists a sequence $\{f_n\} \subset C(X)$ that converges to $f$ in
$L^2(X)$. Then we have
$$
\|\pi_\omega(f_n -f_m)\varphi\|_2^2 = \|(f_n -f_m)\varphi\|_2^2  \leq \|\varphi\|_\infty^2 
\|(f_n -f_m)\|_2^2 \rightarrow 0.
$$
In an analogous way, we can prove that $\|\pi_\omega(\overline{f}_n -\overline{f}_m)\varphi\|_2^2 \rightarrow 0$. 
Therefore the sequence $\pi_\omega(f_n)$ is $t_{s^*}$-Cauchy.

\vspace{4mm} 
(2) 
 Let $X=[0,1]$, $\A_o = C(X)$ and $\A=L^p(X), \; p \geq 1$, as in Example 1. 
Let, moreover, $w \in L^r(X), \; r \geq 1$, and $w>0$.
Let $\omega$ be the linear functional on $C(X)$ defined by
$$ 
\omega (f) = \int_X f(x)\, w(x) \;dx, \quad f \in C(X).
$$ 
The sesquilinear form  $\Omega_o$ associated with $\omega$ is then defined as
$$ 
\Omega_o (f,g)= \int_X f(x)\,\overline{g(x)} \,w(x) \; dx, \quad f,g \in C(X).
$$
Let us discuss the closability of $\Omega_o$ when $p$ varies in $[1,\infty)$.
The whole discussion can be reduced to the cases examined in Example 1, if we take into account
 the following facts:
\begin{itemize}
\item[(i)] $\Omega_o(f,f)= \|fw^{1/2}\|_2^2, \quad \forall f \in C(X)$; 
\item[(ii)] If $\|f_n\|_p \to 0$, then $\|f_n w^{1/2}\|_s \to 0$, where 
 $s^{-1} =p^{-1} + (2r)^{-1}$;
\item[(iii)] $\displaystyle \int_X w(x) \,dx< \infty$.
\end{itemize}
The conclusion is that $\Omega_o$ is bounded if $s \geq 2$, while, for $1\leq s <2$, $\Omega_o$ 
is not bounded, but it is closable.

\vspace{4mm} 
(3) 
We end this section with a nonabelian example.  Let $\A$ be the vector space of all infinite
matrices
$A:=(a_{mn})$ satisfying the condition
$$ 
\|A\|^2:= \sum_{m,n=1}^\infty \frac{1}{m^2 n^2}\;|a_{mn}|^2 < \infty .
$$
$\A$ is a Banach space with respect to this norm. With the ordinary matrix multiplication and
the usual involution $A \mapsto A\ha$, \A\  may also be regarded as a topological
\qxa\  over the *-algebra $\A_o$ of all matrices with a finite number of non-zero
entries in each row and in each column (this is a \qxa\  without unit, since 
$I \in \A \backslash\A_o$).

Let 
$$
\omega(A)= \mbox{tr}(A)= \sum_{m}^\infty a_{mm}, \quad A\in \A_o.
$$
The sesquilinear form $\Omega_o$ associated to $\omega$ is then defined by
$$
\Omega_o(A,B)= \omega(B\ha A)=\sum_{m,n=1}^\infty \overline{b_{mn}} a_{mn},\quad A, B\in \A_o,
$$ 
with $B:=(b_{mn})$.
Let us show that $\Omega_o$ is closable.
Let $\{A_k\}$, with $A_k = (a^k_{mn}),$ be a sequence in $\A_o$ such that 
$\|A_k\|\rightarrow 0$ and 
 $\Omega_o(A_k-A_j,A_k-A_j)$ converges to $0$. We will prove that $\Omega_o(A_k,A_k)
\to 0$. In terms of matrices, the two conditions above read
\be 
\sum_{m,n=1}^\infty \frac{1}{m^2 n^2}\;|a^k_{mn}|^2 \rightarrow 0, 
\quad \mbox{for } k \rightarrow \infty,
\label{one}
\en
and 
\be 
\sum_{m,n=1}^\infty |a^k_{mn}-a^j_{mn}|^2\rightarrow a, \quad \mbox{for } k \rightarrow \infty.
\label{two}
\en
Let $\gB$ denote the vector space of all infinite matrices $A:=(a_{mn})$ satisfying the condition
$$ 
\|A\|_2^2:= \sum_{m,n=1}^\infty |a_{mn}|^2 < \infty .
$$
$\gB$ also is a Banach space with respect to this norm. Then (\ref{two}) means that $\{A_k\}$ is
a Cauchy sequence with respect to $\| .\|_2$ and it converges to a matrix $A\in \gB$. Thus
necessarily 
\be\|A_k\|_2^2=\Omega_0(A_k,A_k)=\sum_{m,n=1}^\infty |a^k_{mn}|^2\to\|A\|_2:=a. 
\label{three}
\en
Now set $M= \sum_{m,n=1}^\infty \frac{1}{m^2 n^2}$. Then (\ref{three}) can be cast in the 
following form 
\be 
\sum_{m,n=1}^\infty\frac{1}{m^2 n^2}\left(m^2n^2 |a^k_{mn}|^2- \frac{a}{M}\right)\rightarrow 0,
\quad \mbox{for } k \rightarrow \infty.
\label{four}
\en
This implies that, for all $m,n \in \BN$,
$$
|a^k_{mn}|^2 - \frac{a}{m^2n^2M}\rightarrow 0, \quad \mbox{for } k \rightarrow \infty.
$$
But from (\ref{two}) we get also that, for all $m,n \in \BN$,  $|a^k_{mn}|^2\rightarrow 0, 
\;\mbox{for } k \rightarrow \infty.$ Hence $a=0$.

The next point is to identify $\A_\Omega$. 
We claim that, in this case, $\D_{\Omega}=\A_\Omega={\gB}$. 
The inclusion ${\gB}\subset\D_{\Omega}$ is easy.
Let now $A=(a_{mn}) \in \D_{\Omega}$; then there exists a sequence $\{A_k\}$, with 
$A_k = (a^k_{mn})$ such that $\|A-A_k\| \rightarrow 0, \;\mbox{for } k \rightarrow \infty$ and
$\Omega_o(A_k,A_k)$ convergent. From $\|A-A_k\| \rightarrow 0$, it follows easily that
$a^k_{mn}$ converges to $a_{mn}$ for each $m,n$. Moreover, $\Omega_o(A_k,A_k)$ converges to
$\Omega(A,A)$ by definition. Finally, we get
$$
\Omega(A,A)= \lim_{k \rightarrow \infty}\sum_{m,n=1}^\infty |a^k_{mn}|^2 =\sum_{m,n=1}^\infty 
|a_{mn}|^2<\infty.
$$
Therefore $\D_{\Omega}\subset{\gB}$. The equality $\D_{\Omega}=\A_\Omega$ is, in this case, obvious.

\section{\hspace{-5mm}. Extensions by sesquilinear forms}

In Section 2, we have studied the extension of representations by Hilbert space operators,
 $\pi(A) \in \LD$. Now we turn to extensions in the space of sesquilinear forms $\LDD.$

\subsection{\hspace{-5mm}. Quasi *-algebras generated by *-representations}

First we consider a \xa\  $\A_o$ on its own and show that a *-representation $\pi_o$ of $\A_o$
  can be used to built up a \qxa\  related to $\pi_o$.

Let $\pi_o$ be a *-representation of the  *-algebra $\A_o$ defined on a certain
domain $\D (\pi) :=\D$, dense subspace of a given Hilbert space \H. This means that the
linear map $\pi_o$, which maps $\A_o$ into $\LpD$, is such that
$\pi_o(A\ha)=\pi_o(A)\ad $ and $\pi_o(AB)=\pi_o(A)\pi_o(B)$ for all $A$ and $B$ in
$\A_o$. 

Let us now endow $\D$ with the
graph topology $t_{{\cal L}\sad} $ generated by the following family of seminorms:
$$
\varphi \in \D \mapsto \| X\varphi \|; \:\: X\in \LpD.
$$
Then we construct the  rigged Hilbert space (RHS) \cite{gelf}:
$$
\D[t_{{\cal L}\sad} ]\subset \H \subset \D'[t_{{\cal L}\sad} '],
$$
where $\D'[t_{{\cal L}\sad} ']$ is the conjugate dual of $\D[t_{{\cal L}\sad}]$
endowed with the strong dual topology.

As usual we will denote with $\LDD$ the vector space of all the linear maps
which are continuous from $\D[t_{{\cal L}\sad}]$ into $\D'[t_{{\cal L}\sad} ']$.
With this construction, we know that $(\LDD, \LpD)$ is a quasi *-algebra
\cite{tra1}.

\noindent
 Many topologies may be introduced in $\LDD$. For instance:

\bei
\item[{\sl (i)}] {\sl  Uniform topology} $\tau ^{\D}$ :
It is defined by the seminorms
$$
A\in \LDD \mapsto \|A\|_{\cal M} = \sup_{\varphi, \psi \in {\cal M}}
|<A\varphi, \psi>|,
$$
where ${\cal M}$ is a bounded subset of $\D$.

\item[{\sl (ii)}] {\sl  Strong topology} $\tau _{s}$ :
For an element $A\in \LDD$ and for a vector $\phi \in \D$ we define the
following seminorms 
$$
\|A\varphi\|_{\cal M} = \sup_{\psi \in {\cal M}} |<A\varphi, \psi>|,
$$
with $\cal M$ as above.

\item[{\sl (iii)}] {\sl Strong* topology} $\tau _{\rm s^*}$ :
In this case the seminorms are 
$$
A\in \LDD \mapsto \max \{\|A\phi\|_{\cal M}, \|A\ad\phi\|_{\cal M}\},
$$
where $\|A\phi\|_{\cal M}$ is defined above.

\item[{\sl (iv)}] {\sl Weak topology} $\tau _{\rm w}$ :
The seminorms are 
$$
A\in \LDD \mapsto |<A\varphi, \psi>|, 
$$
where $\varphi$ and $\psi$ belong to $\D$.
\eni
The involution $A \mapsto A\ha$ is continuous for $\tau ^{\D}, \,\tau _{\rm s^*} $ and
$\tau _{\rm w} $, but of course not for $\tau _{\rm s} $. The multiplication from the
 right by elements from $\LpD$ is continuous for $\tau ^{\D}$ and
$\tau _{\rm w} $, and that from the left is continuous for $\tau ^{\D}, \,\tau _{\rm s} $ and
$\tau _{\rm w} $.
 
Whenever $\pi_o$ is a faithful representation of the *-algebra $\A_o$, we can
introduce on $\A_o$ a topology which is linked to the one introduced in
the representation space. Let us assume, for instance, that $\LDD$ is endowed
with the uniform topology $\tau ^{\D}$. Then, if $\cal M$ is a bounded set in 
$\D[t_{{\cal L}\sad}]$, we define a seminorm on $\A_o$ by
$$
p_{\cal M} (B) \equiv \|\pi_o(B)\|_{\cal M}, \hspace{1cm} B\in \A_o.
$$
Since $\pi_o$ is faithful, this is a separating family of seminorms. Calling 
$\tau ^{\D}_o$ this topology, we can easily conclude that $\A_o[\tau ^{\D}_o]$ is
a locally convex *-algebra.

{\bf Remark.} -- An analogous result holds
if we replace the uniform topology by the weak one. However, $\A_o$ fails to be
a locally convex *-algebra for the corresponding strong and the strong* topologies.
\bigskip \bigskip

Let now $\A$ be the completion of $\A_o$ with respect to $\tau ^{\D}_o$. It
is clear that $(\A[\tau ^{\D}_{o}],\A_o)$ is a topological quasi *-algebra.
By the construction itself, the representation $\pi_o$ is
continuous from $\A_o[\tau ^{\D}_o]$ into $\LDD[\tau ^{\D}]$. As a consequence
we have the following

\begin{prop} -- Let $\pi_o$ be a *-representation of $\A_o$ in $\D$. Then
the following statements hold true:

(i) If $\pi_o$ is faithful, then there exist a locally convex topology $\tau
^{\D}_o$ on $\A_o$ such that $\pi_o$ is continuous from $\A_o[\tau ^{\D}_o]$ 
into $\LDD[\tau ^{\D}]$.

(ii) If $\LDD[\tau ^{\D}]$ is complete, then $\pi_o$ has an extension $\pi$ to
the quasi *-algebra $(\A[\tau ^{\D}_o],\A_o)$, where $\A$ denotes the $\tau
^{\D}_o$-completion of $\A_o$. The map $\pi$ has the following properties:

a) $\pi(A\ha)=\pi(A)\ad, \quad \forall \, A\in \A$;

b) $\pi(AB)=\pi(A)\pi_o(B), \quad \forall\, A\in \A, \; \forall \, B\in \A_o$.
\end{prop}

We will discuss  some examples of this situation below.
Before that, it is worth remarking that the procedure outlined in this section for one faithful 
representation can be easily extended to a faithful family $\Pi_I = \{\pi^o_\alpha,\; \alpha
\in I \}$ of *-representations of $\A_o$. Here {\em faithful} means that, for each non-zero $A \in
\A_o$, there exists $\alpha \in I$ such that $\pi^o_\alpha(A) \neq 0$. Of course, each
$\pi^o_\alpha$ is a *-representation on a domain $\D_\alpha$, that is $\pi^o_\alpha(A)\in
{\cal L}\ad(\D_\alpha)$. Then we can define a  locally convex topology $\tau_{\Pi_I}$  on
$\A_o$ as the weakest locally convex topology such that each $\pi^o_\alpha$ is continuous
from $\A_o$ into
${\cal L}(\D_\alpha, \D'_\alpha )[\tau_{op}]$ and proceed as before to get an obvious
 extension of Proposition 3.1.

\vspace{3mm} 

\noindent{\bf Examples 3.2}
(1)
Let $X:=[0,1]$ and $P$ the selfadjoint operator defined on
$$
D(P)\equiv\left\{f\in L^2(X): f \mbox{ is absolutely  continuous},
f' \in  L^2(X),\,  f(0)=f(1)\right\}
$$
by $$(Pf)(x)\equiv -if'(x).$$ 
Define the domain
$$
\D=\left\{f\in C^\infty(X): \:\: f^{(n)}(0)=f^{(n)}(1), \: \forall \,n \in \BN
\cup\{0\} \right\}.
$$
Then $\D$ coincides with $\D^\infty(P)$, and the topology
$t_{{\cal L}\sad} $ coincides with the topology given by the following family of
seminorms:
$$
\varphi \in \D \mapsto\|\varphi\|^2_k = \|(1+P^2)^k\varphi\|,
\hspace{2cm} k=0,1,2,\ldots
$$
It is easy to check that $\D$ is a *-algebra and that the
multiplication is jointly continuous  \cite{tra1}: for any $k\in \BN$, there exists a 
positive
constant $c_k$ such that 
\be
\|\varphi \, \chi \|_k \leq c_k \|\varphi \|_k \|\ \chi \|_k, \hspace{2cm}
\forall \varphi,\, \chi \in \D.
\label{1}
\en
We can now define a *-representation $\pi_o$ of $\D$ on $\D$ itself by
$$
\pi_o(f)g = fg, \qquad \forall \, f,g \in \D.
$$
This representation is faithful since the function $u(x)=1, \, \forall x \in [0,1],$ belongs
 to $\D$.

Let $\D'$ be the conjugate dual of $\D$ 
with respect to $t_{{\cal L}\sad} $. Using
(\ref{1}), we see that if $\Phi$ is an element of $\D'$ and $f$ is any vector of
$\D$, then $\Phi f \in \D'$, where the product is defined in the following natural way:
\be
<\Phi f, g> \equiv <\Phi,\overline{f} g>, \quad g \in \D.
\label{2}
\en
One has indeed:
$$
|<\Phi f,g>| = |<\Phi,\overline{f} g>|\;  \leq \; c \|\overline{f} g \|_k 
\; \leq \; c_k \| f \|_k \, \| g \|_k .
$$
If we endow $\D'$ with the strong dual topology, then $(\D',\D)$ is a
topological quasi *-algebra \cite{tra2}. In an obvious way we can define a
*-representation of $\D'$:
\be
\pi(\Phi)f = \Phi f, \qquad  \forall \,\Phi \in \D',\: \forall f \in \D.
\label{3}
\en
 In this case $\pi (\Phi) \in
\LDD$, $\pi$ extends the representation $\pi_o$ and it is faithful too. We want to
show that this representation is exactly the one defined in the first
part of this section.

In order to do this, we start by introducing the sets ${\cal B} = \{\pi(\Phi):
\:\: \Phi \in \D'\}$ and ${\cal B}_o = \{\pi_o(f): \:\: f \in \D\}$. We have to
prove that ${\cal B}$ is uniformly complete and that ${\cal B}_o$ is dense in
it. This can be proven easily  by showing that the $\tau_o^\D$-topology on $\D$
is equivalent to the topology induced by $\D'$ on $\D$ itself. Let indeed $\cal M$ be a
bounded subset of $\D[t_{{\cal L}\sad}]$, then we have:
\begin{eqnarray*}
\|\pi(\Phi)\|_{\cal M} &=& \sup_{f,g \in {\cal M}}|<\pi(\Phi)f,g>| \\  &=& 
\sup_{f,g \in {\cal M}}|<\Phi,f^* g>| = \sup_{h \in {\cal M}\cdot {\cal
M}}|<\Phi,h>| = \|\Phi\|_{\cal M\cdot M}.
\end{eqnarray*}
It may be worthwhile to remind that, of course, the joint continuity of the
multiplication implies that the set $\cal M\cdot M$ is bounded.

On the other hand if we consider the norm $\|\Phi\|_{\cal M}$, then we have
\begin{eqnarray*}
\|\Phi\|_{\cal M}&=& \sup_{f \in {\cal M}}|<\Phi,f>| = \sup_{f \in {\cal
M}}|<\pi(\Phi)u,f>|\\ &\leq& \sup_{f,g \in {\cal M}\cup \{u\}}|<\pi(\Phi)g,f>|=
\|\pi(\Phi)\|_{{\cal M}\cup \{u\}},
\end{eqnarray*}
where $u(x)$ is the unit function.
\vspace{3mm} 

(2)
Our second example is that of the CCR algebra on an interval \cite{lasst}. 
The starting point is again the
rigged Hilbert space considered in the previous example, 
$$
\D \subset L^2([0,1]) \subset \D'.
$$
Let now $\A_o$ denote the vector space of all formal polynomials $Q$ in one variable
$p$ with coefficients in $\D$, i.e. $Q= \sum_{k=0}^N\varphi_k p^k$, $N \in \BN$ with 
$\varphi_k \in \D$.

$\A_o$ can be made into an  algebra by introducing a multiplication in the following
way: if $Q_1= \sum_{k=0}^N\varphi_k p^k$ and $Q_2= \sum_{l=0}^M\psi_l p^l$,
then we put
$$
Q_1Q_2 := \sum_{k=0}^N \sum_{l=0}^M \varphi_k \left(\sum_{r=0}^k(-i)^r
\left(
\begin{array}{c}
k  \\ 
r \\ 
\end{array}
\right)
\frac{d^r\psi_l}{dx^r}p^{k-r+l}\right).
$$
An involution can also be introduced easily in $\A_o$ by means of the following
formula:
$$
\left( \sum_{k=0}^N \varphi_k p^k\right)^* := \sum_{k=0}^N 
 \sum_{r=0}^k(-i)^r \left(
\begin{array}{c}
k  \\ 
r \\ 
\end{array}
\right)
\frac{d^r\overline{\varphi}_k}{dx^r}p^{k-r}.
$$
With this definition $\A_o$ is a *-algebra. Let now $P$ be the operator defined
in Example 1. As already mentioned, in this case $\D = \D^\infty(P)$ and the topology
$t_{{\cal L}\sad} $ coincides with the graph topology defined by $P$. Of course
$P\in \LpD$, so that it admits a unique extension, again indicated with the
same symbol, to $\D'$, defined by
$$
<P\Phi, \varphi> = < \Phi, P\varphi> \hspace{2cm} \varphi \in \D, \: \Phi \in
\D'.
$$
Let $\widehat \Phi$ be the multiplication
operator defined, for $\Phi \in \D'$,  as in (\ref{2}). Then $ \widehat \Phi P^k \in \LDD$ 
and, in
particular, if $\Phi \in \D$, then $\widehat \Phi P^k $ belongs to $\LpD$. This allows us to
define a representation $\pi_o$ of $\A_o$ on $\D$ in the following way:

$$ \pi_o: Q= \sum_{k=0}^N\varphi_k p^k \in \A_o \mapsto
\pi_o(Q)= \sum_{k=0}^N \hat \varphi_k
P^k \in   \LpD.$$

The representation $\pi_o$ is faithful; indeed if $\pi_o(Q)=0$, then
$\sum_{k=0}^M\hat\varphi_k P^k\psi=0$, for all $\psi\in \D$. Now, for $\psi(x)= u(x) =1,\, 
\forall x \in [0,1]$, we get $\varphi_0=0$; the choice of $\psi(x)=x,\, \forall x \in [0,1]$,
implies that $\varphi_1=0$, and so on. Thus we may have $\pi_o(Q)=0$ only if all coefficients
$\varphi_k$ are zero. This in turn implies that $Q=0$.

Let us now endow $\A_o$ with the topology $\tau_o^\D$ defined by the seminorms
$$
Q\in \A_o \mapsto \|\pi_o(Q)\|_{\cal M},
$$
where ${\cal M}$ is a bounded subset in $\D$.  It is shown in \cite{tra1} that
the completion $\widehat{ \A}_o$ of $\A_o$ with respect to this topology contains the
following space
$$
{\A} = \left\{ \sum_{k=0}^N F_k p^k,  \quad F_k \in \D'\right\}.
$$
Incidentally we observe that $(\widehat{\A}_o, \A_o)$ is a quasi *-algebra. The
representation $\pi_o$ can be extended to the whole $\widehat {\A}_o$. This extension is 
clearly given by: 
$$ 
\pi \left( \sum_{k=0}^N F_k p^k\right)= \sum_{k=0}^N {\widehat F}_k P^k.
$$ 
It is easy to check (by means of the same technique as in the previous example)
that $\pi$ coincides
again with the one discussed in the first part of  this section.

\subsection{\hspace{-5mm}. Sesquilinear form extensions within a given \qxa}

We have proven in the previous section  that a faithful *-representation $\pi_o$
 of a \xa\ $\A_o$ generates a topological \qxa\, to which $\pi_o$ can be extended.
Now we consider the problem of the extension of a *-representation when the topological \qxa\  
is given a priori.

Let $(\A[\tau], \A_o)$ be a topological \qxa\ and $\pi_o$ a *-representation of $\A_o$ on the 
domain $\D (\pi_o):= \D$. As we have seen in the introduction, one has
\be
\LpD \; \subset \; \LD \; \subset \; \LDD.
\en
Thus another possibility for extending  $\pi_o$ is to impose closability in $ \LDD$
instead of $\LD$. This requires that we consider the various topologies described in Section
3.1.  Thus we  assume that $\LDD$ is endowed with $\tau_{op}$, where $\tau_{op}$ stands for any
of the topologies $\tau^{\D}$, $\tau_{\rm s^*}$, $\tau_{\rm s}$, $\tau_{\rm w}$.

\setcounter{theorem}{2}
\begin{defin} -- 
We say that $\pi_o$ is $\tau_{op}$-extendible if it is closable as a linear map from $\A _o
[\tau]$  to $\LDD [\tau_{op}]$. This means that, for any net $\{X_\alpha\} \subset \A _o $ such
that 
$ X_\alpha \stackrel{\tau}{\longrightarrow} 0 {\mbox{ and }} \pi_o(X_\alpha) \stackrel{\tau_{op}}
{\longrightarrow} Y\in \LDD,$
it follows that  $Y=0$.
\label{main2}
\end{defin}
It is clear that the four notions of extendibility we have given compare in the following way:

$\tau^{\D}$-extendible $\;\Rightarrow\;$ $\tau_{\rm s*}$-extendible $\;\Rightarrow\;$
$\tau_{\rm s}$-extendible $\;\Rightarrow\;$$\tau_{\rm w}$-extendible.

\noindent 
If $\pi_o$ is $\tau_{op}$-extendible, then we put
$$
\A(\pi,\tau_{op})=
\left\{ X \in \A \mid \exists \{X_\alpha \}\subset \A _o: X_\alpha 
\stackrel{\tau}{\longrightarrow}
 X
{\mbox{ and }} \pi_o(X_\alpha)\,\mbox{is}\, \tau_{op}\mbox{ -convergent in } \LDD \right\}. 
\label{dom}
$$
For $X \in \A(\pi,\tau_{op})$, we set $\pi (X) =  \tau_{op}\mbox{-}\lim\pi_o(X_\alpha)$. Thus  
$\pi$ is well-defined and extends $\pi_o$.

\belem
--
If $\tau_{op} \neq \tau_{\rm s}$, then $X \in \A(\pi,\tau_{op})$ implies $X\ha \in 
\A(\pi,\tau_{op})$. 

If $\tau_{op} = \tau^{\D}$ or $\tau_{op} = \tau_{\rm w}$, then $X \in \A(\pi,\tau_{op}),\, 
A \in \A_o$ imply $AX, XA \in \A(\pi,\tau_{op}).$
\end{lemma}
\proof 
The first statement depends on the continuity of the involution; the second on the continuity of 
the multiplications, as discussed in Section 3.1.
\enproof

Thus we get two \qxa s over $\A_o$.

\begin{prop}
--
$\A(\pi,\tau^{\D})$ and $\A(\pi,\tau_{\rm w})$ are \qxa s over $\A_o$.
\end{prop}

Of course, what we want are {\em topological} \qxa s, and this requires some additional input.
Let $p_\alpha$ be a (directed) family of seminorms generating the topology $\tau$ of $\A$ and
 $q_\beta$ a (directed) family of seminorms generating $\tau_{op}$, where $\tau_{op} =
\tau^{\D}$ or $\tau_{op} = \tau_{\rm w}$. Then we can define a new topology $\eta_{op}$ on
$\A(\pi,\tau_{op})$ by the family of seminorms 
$$ \eta_{\alpha, \beta}(X) = p_\alpha (X) + q_\beta (\pi(X)).$$

By the construction itself it follows that
\begin{prop}
--
$\A(\pi,\tau_{op})[\eta_{op}]$ is a topological  \qxa\ over $\A_o$. If both $\A [\tau]$ and
$\LDD [\tau_{op}]$ are complete, then $\A(\pi,\tau_{op})[\eta_{op}]$ is complete.
\end{prop}

\noindent
 If $\LDD$ is $\tau_{op}$-complete, then $\A(\pi,\tau_{op})$  can be rewritten in the following way
\be
\A(\pi,\tau_{op})= \left\{ X \in \A \mid \exists \{X_\alpha \}\subset \A _o: X_\alpha
\stackrel{\tau}{\longrightarrow} X \,,\, \pi_o(X_\alpha) \mbox{ is a ${\tau_{op}}$-Cauchy
net}\right\}.
 \label{dom2}
\en
It turns out that in several examples this domain \D\ has a special form. This happens, 
for instance, when there exists a self-adjoint operator $H$ in Hilbert space \H\ such that 
$\D = \D^\infty (H)$. 

In this case the topology $t_{{\cal L}\sad}$ coincides with the topology defined by the
seminorms
$$ 
\phi \in \D \mapsto \|\phi\|_n =\|H^n \phi\|, \quad n \in \BN\cup \{0\} .
$$
Without loss of generality, we may assume that $H \geq 1$; in this case we have 
$$ 
\|\phi\|_n \leq \|\phi\|_{n+1}, \quad \forall n \in \BN\cup \{0\}.
$$
Furthermore, $\D$ is a reflexive Fr{\'e}chet space, $\D '$ is complete for the strong dual
topology and, when endowed with the uniform topology $\tau^{\D}$, $(\LDD, \LpD)$ is a topological
\qxa\ and $\LDD$ is $\tau^{\D}$-complete \cite{lass1,tra1}.

 The same statement is true if $\tau_{op}=\tau_{\rm s^*}$. Indeed,
\begin{prop}
--
If $\D =\D^\infty (H)$, then $\LDD [\tau_{\rm s^*}]$ is a complete locally convex space with 
ontinuous involution.
\end{prop}
\proof
Let $\{A_\alpha \}$ be a $\tau_{\rm s^*}$-Cauchy net in $\LDD$. Then by the definition it 
follows
that, for each $\phi \in \D$, $\{A_\alpha \phi\}$ and
$\{A\ad_\alpha \phi\}$ are Cauchy nets in $\D '$ with the strong dual topology. Since $\D '$ is 
complete, there exist $\Phi, \Psi \in \D '$ such that $A_\alpha \phi \rightarrow \Phi$ and
$A\ad_\alpha \phi \rightarrow \Psi$. Set $A\phi =\Phi$ and $B\phi=\Psi$. Is is easy to see that 
the
following equality is fulfilled
$$ 
<A\phi,\psi>=<\phi,B\psi>, \quad \forall \, \phi, \psi \in \D.
$$ 
By the reflexivity of $\D$, this implies that $ A\in \LDD$ and $B=A\ad$. The convergence of 
$A_\alpha$ to $A$ is clear. 
\enproof

However, $\LDD [\tau_{\rm s^*}]$ is  not a topological  \qxa\ on $\LpD$ since the multiplication
 may fail to be continuous for $\tau_{\rm s^*}$, as we mentioned already.

\vspace{4mm}
\noindent {\bf Examples 3.8}.

\noindent{\sl (1) Representations of the polynomial algebra}
\medskip

Let $\A_o$ be the *-algebra of all polynomials in one real variable with complex coefficients 
and let $A$ be a self-adjoint operator in \H\ , $\D =\D^\infty(A)$ and $\pi_o$ the following
representation of $\A_o$:
$$ \A_o \ni \sum_{k=1}^n \lambda_k x^k \mapsto \sum_{k=1}^n \lambda_k A^k \in \LpD.$$

Let $\A = L^1(\BR, e^{-x^2/2}dx)$. Then $(\A, \A_o)$ is a topological *-algebra with respect to 
the $L^1$-norm. 
We will now discuss the $\tau^{\D}$ extendibility of $\pi_o$.

Let $p_n(x)$ be a sequence of polynomials in $\A_o$ such that

$$ \int_{\BR} |p_n(x)|e^{-x^2/2}dx \to 0 \mbox{ when } n \to \infty.$$

As is well known, when $\D =\D^\infty(A)$, the topology $\tau^{\D}$  can be described by the 
seminorms
$$ 
X \mapsto \|f(A)Xf(A)\|, \quad X \in \LDD, 
$$
where $f(x)$ runs over the set $\cal F$ of all bounded continuous functions on $[0,\infty)$ such 
that $ \sup_{x \in \BR^+} x^k f(x) < \infty, \; \forall k \in \BN$.
Let us now assume that 
$$ 
\pi_o(p_n(x))= p_n(A) \stackrel{\tau^{\D}}{\longrightarrow} Y,
$$
which in the present case means that 
$$
\|f(A)(p_n(A)-Y)f(A)\|\to 0.
$$
By definition, $\pi_o$ is $\tau^{\D}$-extendible if this condition  implies $Y=0$. 
If $A$ is the multiplication operator by $x$ on $L^2(\BR)$ and $p_n(x)$ converges to zero 
in $L^1(\BR, e^{-x^2/2}dx)$, then, since $\mu(\BR)<\infty$, we can find a subsequence 
$p_{n_k}(x)$
which converges to $0$ $\mu$-a.e. This fact easily implies that, if $\|f^2(x)p_n(x)\|$ converges,
then its limit is $0$ and so $Y=0$. Therefore, in this case, $\pi_o$ is $\tau^{\D}$-extendible.
However, this does not seem to be necessarily the case for an arbitrary self-adjoint operator $A$,
so the question remains open in general.

\bigskip

\noindent{\sl (2) Multiplication operators}
\medskip

Let $X=[0,1]$, $\A=L^1(X)$ and $\A_o = C(X)$, the C*-algebra of continuous functions on $X$.
We define a representation $\pi_o$ of $C(X)$ on $\D= L^p(X)$ $(p>2)$, considered as a dense 
subspace of $L^2(X)$, by
$$ \pi_o(f)\phi = f\phi, \quad \phi \in L^p(X).$$
The first step in our construction consists in getting sufficient information on ${\cal L}\ad
 (L^p(X))$. We claim that ${\cal L}\ad (L^p(X))$ consists only of bounded operators in $L^2(X)$.
Indeed, if $T \in {\cal L}\ad (L^p(X))$, then $T$ is closable in $L^2(X)$. Assume now that $\|f_n
\|_p \rightarrow 0$ and $\|Tf_n - g\|_p\rightarrow 0$, then $\|f_n \|_2 \rightarrow 0$ and $\|Tf_n
- g\|_2\rightarrow 0$. Thus $g=0$. Therefore $T$ is  closed and everywhere defined in $L^p(X)$ and
so it is bounded in $L^p(X)$. Analogously $T\ha$ is bounded in $L^{\overline{p}}(X)$ with
$p^{-1}+\overline{p}^{-1}=1$. Exchanging the roles of $T$ and $T\ha$, it turns out that $T$ is
bounded in both $L^p(X)$ and $L^{\overline{p}}(X)$ and therefore (by interpolation) in any $L^r(X)$
with ${\overline{p}}\leq r \leq p$. We conclude that $T$ is bounded in $L^2(X)$.

Hence, the graph topology on $\D = L^p(X)$ coincides with the $L^2$-norm on $\D$; thus $\D '= 
\H =L^2(X)$. Therefore $\LDD = {\cal B}(\D,\H)$, the set of all bounded operators from \D\ into \H\
(which is isomorphic to ${\cal B}(\H)$). We remark that ${\cal B}(\D,\H)$ is complete for
$\tau^{\D}$ (which coincides with the uniform topology of ${\cal B}(\H)$), but not for $\tau_{\rm
s^*}$. In fact, the $\tau_{\rm s^*}$-completion of ${\cal B}(\D,\H)$ is
$\LD$ (incidentally, this shows that $L^p(X) \neq \D^\infty (A)$ for any self-adjoint operator $A$).
We show now that $\pi_o$ is {\em closed} as a densely defined linear map from $C(X) \subset L^1(X)$ into ${\cal B}(\D,\H)$.
First, $\pi_o$, like any representation of a C*-algebra, is continuous from $C(X)$, with its
 C*-norm, into ${\cal B}(\H)$. Moreover, $\pi_o$  is faithful and so it is isometric, i.e.,
$$
\|\pi_o(f)\| = \|f\|_\infty, \quad f \in C(X).
$$
 Let now $\{f_n\}$ be a sequence in $C(X)$ with the properties:
$$ 
\|f_n -f\|_1\to 0 \mbox{ and } \pi_o(f_n) \to Y \mbox{ uniformly }.
$$
Then
$$ 
\|f_n -f_m\|_\infty = \|\pi_o(f_n) - \pi_o(f_m)\|\to 0. 
$$
Hence, it follows from the completeness of $C(X)$  that $f \in C(X)$ and $Y=\pi_o(f)$.
Thus $\pi_o$ does not admit extensions by closure to $\A$.

\vspace{3mm}
The situation changes drastically if we weaken somewhat the definition of $\tau_{op}$-extendible
representation.

Let $\pi_o$ be $\tau_{op}$-extendible in the sense of Definition \ref{main2} and let
\be
\widetilde{\A}(\pi,\tau_{op})= 
\left\{ X \in \A \mid \exists \{X_\alpha \}\subset \A _o: X_\alpha \stackrel{\tau}{\longrightarrow}
 X {\mbox{ and }} \pi_o(X_\alpha) \mbox{ is a ${\tau^\D}$-Cauchy net}\right\}.
 \label{dom3}
\en
Let ${\CLDD}$ denote the completion of $\LDD[\tau_{op}]$. Then for $X \in 
\widetilde{\A}(\pi,\tau_{op})$ we set
$$
\pi(X)=\tau_{op}\mbox{-}\lim_{\alpha}\pi_o(X_\alpha)\in {\CLDD}.
$$
Clearly, if $\LDD$ is $\tau_{op}$-complete, then
 $\widetilde{\A}(\pi,\tau_{op}) = \A(\pi,\tau_{op})$.

Let us come back to Example 2. As we have seen before, in this case, $\LDD = {\cal B}(\D,\H)$.
 Consider on it the topology $\tau_{\rm s^*}$, defined by the seminorms
$$ 
A\in {\cal B}(\D,\H) \mapsto \max\{\|Af\|,\|A\ad f\|\}, \quad f \in \D.
$$
${\cal B}(\D,\H)$ is not complete in this topology, its completion being $\LD$, as already 
noticed.

Now we show that $\pi_o$ is $\tau_{\rm s^*}$-extendible. Indeed, let $\|f_n\|_1 \to 0$ and
 $\pi_o(f_n)\stackrel{\tau_{\rm s^*}}{\longrightarrow} Y$. Then in particular $f_n\phi \to Y\phi$,
in the $L^2$-norm for each $\phi \in \D = L^p(X)$. 

Now, since $\|f_n\|_1 \to 0$, there exists a subsequence $\{f_{n_k}\}$ converging a.e. to $0$. 
Similarly, since $f_{n_k}\phi \to Y\phi$, in the $L^2$-norm, for each $\phi \in \D = L^p(X)$, we
can find a sub-subsequence $\{f_{n_{k_l}}\}$ such that $f_{n_{k_l}}\phi$ converges a.e. to $Y\phi$.
Then, necessarily, $Y\phi = 0$ a.e. and so $Y=0$.

Let us now determine the $\tau_{\rm s^*}$-extension of $\pi_o$.
We will show that $\widetilde{\A}(\pi,\tau_{\rm s^*})= L^s(X)$ where $s =\frac{2p}{p-2}$.
First, we prove that $L^s(X)\subset \widetilde{\A}(\pi,\tau_{\rm s^*})$. Indeed, if $f \in 
L^s(X)$ then there exists a sequence $\{f_k\} \subset C(X)$ such that $\|f_k - f\|_s \rightarrow
0$; then if $\phi \in L^p(X)$ we get
$$\|\pi_o(f_k)\phi-\pi_o(f_l)\phi\|_2 =\|f_k\phi-f_l\phi\|_2 \leq \|f_k -f_l\|_s \|\phi\|_p
\rightarrow 0.$$
In the same way,
$$
\|\pi_o(f_k)\ad\phi-\pi_o(f_l)\ad\phi\|_2 \rightarrow 0.
$$
Then $\tau_{\rm s^*}\mbox{-}\lim_{k \rightarrow \infty} \pi_o(f_k)$ exists in $\LD$. We call 

Finally, since \cite{btellepi}
\be
\left\{f \in L^1(X): f\phi \in L^2(X), \forall \phi \in L^p(X)\right\}= L^s(X),
\quad s = \frac{2p}{p-2},
\en
we conclude that $\widetilde{\A}(\pi,\tau_{\rm s^*})= L^s(X)$.                                                                                                                                          

\section*{Acknowledgments}

This work was performed in the
Institut de Physique Th\'eorique, Universit\'e Catholique de Louvain  and 
the Istituto di Fisica, Universit\`a di Palermo. We thank both institutions
for their hospitality, as well as travel grants from CGRI, Communaut\'e
Fran\c caise de Belgique, Belgium and Ministero degli Affari Esteri, Italia. 
F.B. thanks M.U.R.S.T. (Italy) for financial support.


\begin{thebibliography}{99}

\bibitem{jafglim}A. Jaffe and J. Glimm, {\em Quantum Physics, A Functional Integral Point of
             View}, Springer-Verlag, Berlin et al., 1981

\bibitem{brat}	O. Bratteli and D .Robinson, {\em Operator Algebras and  Quantum 
                 Statistical Mechanics I, II}, Springer-Verlag, Berlin et al., 1979

\bibitem{thir} W. Thirring and A. Wehrl, On the mathematical structure of the B.C.S. model,
                {\em Commun.Math. Phys.} {\bf 4} (1967) 303-314
   
\bibitem{lass1} G. Lassner, Topological algebras and their applications  in Quantum  
      Statistics, {\em Wiss. Z. KMU-Leipzig, Math.-Naturwiss. R.} {\bf 30}  (1981) 572-595

\bibitem{lass2} G. Lassner, { Algebras of unbounded operators and   quantum dynamics,}  
                  {\em Physica}  {\bf 124~A} (1984) 471-480 

\bibitem{lasst} G. Lassner, G.A. Lassner and C. Trapani, {Canonical commutation  
         relations on the interval,} {\em J. Math. Phys.} {\bf 28} (1987) 174-177

\bibitem{bt1} F. Bagarello and C. Trapani, `Almost' mean field Ising model: An algebraic
             approach, {\em J. Stat. Phys.} {\bf  65 } (1991) 469-482

\bibitem{bt2} F. Bagarello and C. Trapani, A note on the algebraic approach 
    to the almost mean field Heisenberg model,
               {\em Nuovo Cim.}  {\bf 108B} (1992) 849-866

 \bibitem{schm} K. Schm\"udgen, {\em Unbounded Operator Algebras and   
       Representation Theory}, Akademie-Verlag, Berlin 1990

\bibitem{ait12} J-P. Antoine, A. Inoue and C. Trapani, Partial *-algebras of 
       closable operators. I. The basic theory and the abelian case.
    II. States and representations of partial *-algebras, 
     {\em  Publ. RIMS, Kyoto Univ.} {\bf 26} (1990) 359-395, {\bf 27} (1991) 399-430

\bibitem{poprev} J-P. Antoine, A. Inoue and C. Trapani, 
     Partial *-algebras of closable operators: A review, 
     {\em  Reviews Math. Phys.} {\bf 8} (1996) 1-42. 

\bibitem{tra1} C. Trapani,  Quasi*-algebras of operators and their applications,
     {\em  Reviews Math. Phys.} {\bf 7} (1995)  1303-1332

\bibitem{boryng} H.J. Borchers and J. Yngvason, On the algebra of field operators. The weak
           commutant and integral decompositions of states, 
                {\em Commun.Math. Phys.} {\bf 42} (1975) 231-252

\bibitem{gelf} I.M. Gelfand and N.Ya. Vilenkin, {\em Generalized Functions}, Vol.IV, 
              Academic Press, New York 1964

\bibitem{tra3} C. Trapani,  States and derivations on quasi*-algebras,
     {\em  J. Math. Phys.} {\bf 29} (1988)  1885-1890

\bibitem{pow}R.T. Powers, Self-adjoint algebras of unbounded operators, 
               {\em Commun.Math. Phys.} {\bf 21} (1971) 85-124

\bibitem{kato} T.Kato, {\em Perturbation theory for linear operators}, Springer-Verlag, 
       Berlin et al., 1966 

\bibitem{weights} J-P. Antoine, Y. Soulet and C. Trapani, Weights on partial
            *-algebras,  {\em  J. Math. Anal. Appl.} {\bf 192} (1995) 920-941

\bibitem{tra2} A. Russo and C. Trapani, Quasi*-algebras and multiplication of distributions,
         preprint No.21, Dip. Mat. Appl., U. Palermo, 1996 

\bibitem{btellepi} F. Bagarello and C. Trapani, $L^p$-spaces as quasi*-algebras,
        {\em  J. Math. Anal. Appl.} {\bf 197} (1996) 810-824   



\end{thebibliography}
\end{document}